# Guidelines for Material Design in Semitransparent Organic Solar Cells

Karen Forberich[1,2], Alessandro Troisi[3], Chao Liu[1,2], Michael Wagner[1], Christoph J. Brabec[1,2], Hans – Joachim Egelhaaf[1,2]

*[1]Forschungszentrum Jülich GmbH, Helmholtz Institute Erlangen-Nürnberg for Renewable Energy (HI ERN), Dept. of High Throughput Methods in Photovoltaics, Erlangen, Germany*
*[2]Friedrich-Alexander-Universität Erlangen-Nürnberg, Materials for Electronics and Energy Technology (i-MEET), Erlangen, Germany*
[3]Dept. of Chemistry, University of Liverpool, United Kingdom

*k.forberich@fz-juelich.de*

**Abstract**

Organic solar cells (OSCs) are uniquely suited for semitransparent applications due to their adjustable absorption spectrum. However, most high-performance semitransparent cells reported to date are based on materials that have shown high power conversion efficiency for opaque devices. We therefore present a model to assess the optimum efficiency and transparency for a specific donor and acceptor band gap. The absorption characteristics of both donor and acceptor are modeled with spectral data of typical absorber materials from the literature which are adjusted to achieve the desired band gap value. The results show three distinct regions of high light utilization efficiency if the photopic curve is employed as a weighting function (corresponding to window applications), and a broad maximum for the plant action spectrum as a weighting function (corresponding to greenhouse applications). When comparing these findings to reported experimental values, it is evident that the band gaps of the materials used for the experimental studies do not correspond to the maxima identified by our simulation model. The analysis of the energy levels of molecules recorded in the literature confirms that all band gaps and therefore all LUE maxima are chemically feasible so that the performance of semitransparent OSCs can be further improved by designing materials with optimized absorption spectra.

**Introduction**

OSCs are an emerging photovoltaics (PV) technology that has seen a steep increase in efficiency over the last five years, with record efficiencies now over 19% [1-3] due to the development of new photon-harvesting materials [4]. Unlike conventional PV technologies such as silicon, OSCs can be straightforwardly fabricated as semitransparent devices in which both the spectrum and the transmittance values can be customized by varying the composition and the thickness of the absorber layer. Building windows and greenhouses are therefore regarded as the most promising applications of OSCs where they can be commercially viable and compete with existing technologies [5,6].

Most reported semitransparent devices have been fabricated with absorber materials that have been developed to yield high efficiency in opaque devices (compare Figure S2 and Tables S1-S3). These absorbers were thus designed to absorb as much light as possible over the whole solar spectrum and convert it into free charges, whereas semitransparent applications require vanishing or very low absorption over the visible region for building integration or the photosynthetically active region (PAR) for agrivoltaics. Accordingly, two parameters have to be optimized simultaneously, namely the power conversion efficiency (PCE) and the transparency, which is quantified by the average visible transmittance (AVT) or average photosynthetically active region transmittance (APART). This makes the selection of suitable absorbers much more complex, as changing the absorption spectra not only affects the transmittance but also the efficiency, due to concomitantly altered short-circuit current ($J_{sc}$) and open-circuit voltage ($V_{oc}$). In this paper, we therefore present a study that is based on the model by Scharber et al. [7,8], with the aim to determine the optimum energy levels for both donor and acceptor and thus provide guidelines for the synthesis of materials specifically tailored for semitransparent devices.

While several studies have assumed idealized rectangular absorption spectra with adjustable boundaries [9-11], and other studies have utilized absorption spectra of reported materials [5,6], we choose to model the absorption of the active layer based on the spectra of exemplary materials from the literature in order to achieve a more realistic description of absorbed and transmitted photons and thus $J_{sc}$ and AVT or APART. These spectra are shifted in energy to account for a variation in the band gap. For the sake of general validity, we have to make assumptions about the input parameters such as fill factor (FF), internal quantum efficiency (IQE), and $V_{oc}$. For instance, the $V_{oc}$ is determined by the energy level difference between the LUMO of the acceptor and the HOMO of the donor, and FF and IQE are set to constant values that is representative of well-working OSCs. The purpose of the model is to demonstrate the potential of a certain material combination, not to exactly reproduce values reported in the literature. After the description of the model and discussing the results in terms of PCE and AVT, we compare the predictions of the model with the values for the best semitransparent devices found in literature, showing that absorbers that have been utilized up to the present do not possess the optimized band gap values. To demonstrate the versatility of our model, we also present the results obtained when using the PAR as the weighting function. Finally, we perform a search of molecules with energy levels available in the literature to assess the chemical feasibility of the optimum band gap values found in our model.

**Description of the model**

Our model is valid for bulk-heterojunction (BHJ) solar cells in which the absorber layer is a mixture of a donor and an acceptor, where charge generation happens at the donor-acceptor interface in the whole absorber layer. The general findings can also be transferred to bilayer solar cells in which acceptor films are deposited on top of donor films and charge generation only happens at the interface if the assumptions for $V_{oc}$ loss, FF are IQE are modified based on experimental values for the best bilayer devices. The calculation is based on the Scharber model for OSCs in which the $V_{oc}$ is calculated from the energy difference between the LUMO of the acceptor and the HOMO of the donor, minus an empirical offset of 0.3 eV [7]:

$$V_{oc} = \frac{1}{e}\left(LUMO_{acceptor} - HOMO_{donor} - 0.3\ eV\right)$$

The energy levels of both donor and acceptor are set as variables, and a minimum energy level difference of 0.2 eV for both the HOMO and the LUMO offset is assumed as necessary for charge transfer (Figure 1a). Even though well-working OSCs with a smaller energy offset have been reported, this offset value of 0.2eV is chosen as it should guarantee efficient charge generation for all material systems [37] and therefore lead to general conclusions.

To calculate the $J_{sc}$, we first assume that the absorption spectra are independent of each other and calculate the optical density (OD) as the sum of optical densities of donor and acceptor according to the donor:acceptor ratio $r_{DA}$:

$$OD_{total}(hv) = OD_{donor}(hv) + \left(\frac{1}{r_{DA}}\right) * OD_{acceptor}(hv)$$

Please note that in this formula $r_{DA}$ describes the peak ratio in the absorbance spectra, which might be different from the weight ratio, depending on the specific material systems and the processing conditions. For instance, in the case of PM6:Y6, in [42] a donor:acceptor weight ratio of 1:1.2 leads to a peak ratio of 1:1, whereas in [43] a weight ratio of 1:1.5 leads to a peak ratio of 1:1.

The obtained OD spectrum is normalized and multiplied with the maximum OD value $OD_{max}$, which represents the active layer thickness:

$$OD_{total}(hv) = \frac{OD_{total}(hv)}{\max(OD_{total}(hv))} * OD_{max}$$

After converting the OD to wavelength-dependent values, the transmittance T is then calculated according to the definition of OD, minus an optical loss that accounts for reflection and parasitic absorption in the non-absorbing layers (such as electrodes and charge transport layers):

$$T(\lambda) = 10^{-OD_{total}(\lambda)} - optical\ loss$$

With this definition, OD refers only to the active layer, the absorption of all the other layers is incorporated into the optical loss. $OD_{max}$ is the maximum value of the absorption and can belong either to the donor or the acceptor peak, depending on the ratio.

Finally, the external quantum efficiency (EQE) is determined from the transmittance as

$$EQE\ (\lambda) = (1 - T(\lambda)) * IQE,$$

where the IQE stands for internal quantum efficiency, and the $J_{sc}$ is obtained from the integration over the product of the EQE and incident illumination spectrum, in this case, the AM1.5G solar spectrum:

$$J_{sc} = \frac{e}{hc} \int EQE(\lambda) * \lambda * \Gamma_{AM1.5G}(\lambda) d\lambda$$

where e is the elementary charge, h is Planck's constant, c is the speed of light, $\lambda$ is the wavelength, and $\Gamma_{AM1.5G}(\lambda)$ is typically given in units of W m$^{-2}$ nm$^{-1}$. If not specified otherwise, the optical loss is assumed as 15%, the IQE as 90% and the fill factor as 70%, representing well-working optimized OSCs.

With respect to the semitransparent properties, the average visible transmittance (AVT) is taken as the integration over transmittance weighted with the photopic curve $P(\lambda)$:

$$AVT = \frac{\int T(\lambda) * P(\lambda) * \Gamma_{AM1.5G}(\lambda) d\lambda}{\int P(\lambda) * \Gamma_{AM1.5G}(\lambda) d\lambda},$$

and the light utilization efficiency (LUE) is used as a figure of merit to assess the potential of the solar cell to provide both high PCE and high AVT [12]:

$$LUE = PCE * AVT.$$

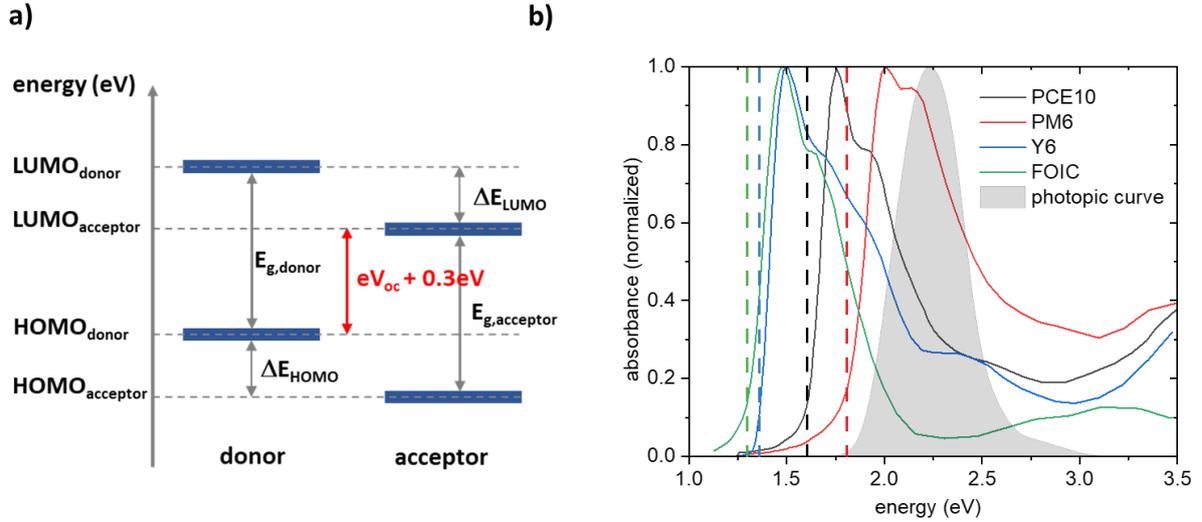

*Figure 1: a) Schematic of the energy levels used to calculate the $V_{oc}$, b) Model absorbance spectra used for the donors PCE10 [13] and PM6 [14] and the acceptors Y6 [14] and FOIC [15]. The band gap values are determined by the intersection point of the rising flank of the spectrum with the wavelength axis. In particular, the values were set as $E_g$ = 1.82 eV or $\lambda_g$ = 680 nm for PM6, $E_g$ = 1.6 eV or $\lambda_g$ = 775 nm for PCE10, $E_g$ = 1.36 eV or $\lambda_g$ = 910 nm for Y6 and $E_g$ = 1.3 eV or $\lambda_g$ = 950 nm for FOIC.*

To model the absorbance spectra for varying band gap ($E_g$), we take typical absorbance spectra of thin films from the literature and shift them on the energy axis by a certain value (either positive or negative) to model a different band gap. The reference value of $E_g$ for a certain model spectrum is assigned based on the intersection of the low energy absorption edge with the wavelength axis (see Figure 1b). Given the various methods employed in literature for measuring the energy levels of absorber materials, each associated with distinct systematic errors [16], we have chosen this approach in which we determine the band gap values from the spectra to have a consistent description. These band gaps do not enter the calculation of the $V_{oc}$ as such, and it is important to note that the spectra are not used to represent a material with all its properties such as HOMO and LUMO values, but only representative absorption spectra.

**Results and Discussion**

To determine the parameters for optimum PCE and AVT, we first vary the donor band gap and HOMO level offset, since the donor typically absorbs in the visible region (see Fig. 1b), determining AVT, and minimizing energy level offset is a known design principle for achieving high PCE. Furthermore, we choose a fixed acceptor band gap of 1.3 eV as this value aligns with the typical band gap value found in most recent acceptors. With this band gap, the absorption spectra lie predominantly outside the visible region (compare spectrum of FOIC in Figure 1b). Furthermore, we use $OD_{max} = 0.7$ and $r_{DA} = 1$ as default values, and the model spectra of PCE10 and FOIC. The resulting plots of PCE, AVT, LUE, $J_{sc}$ and $V_{oc}$ are shown in Figure 2.

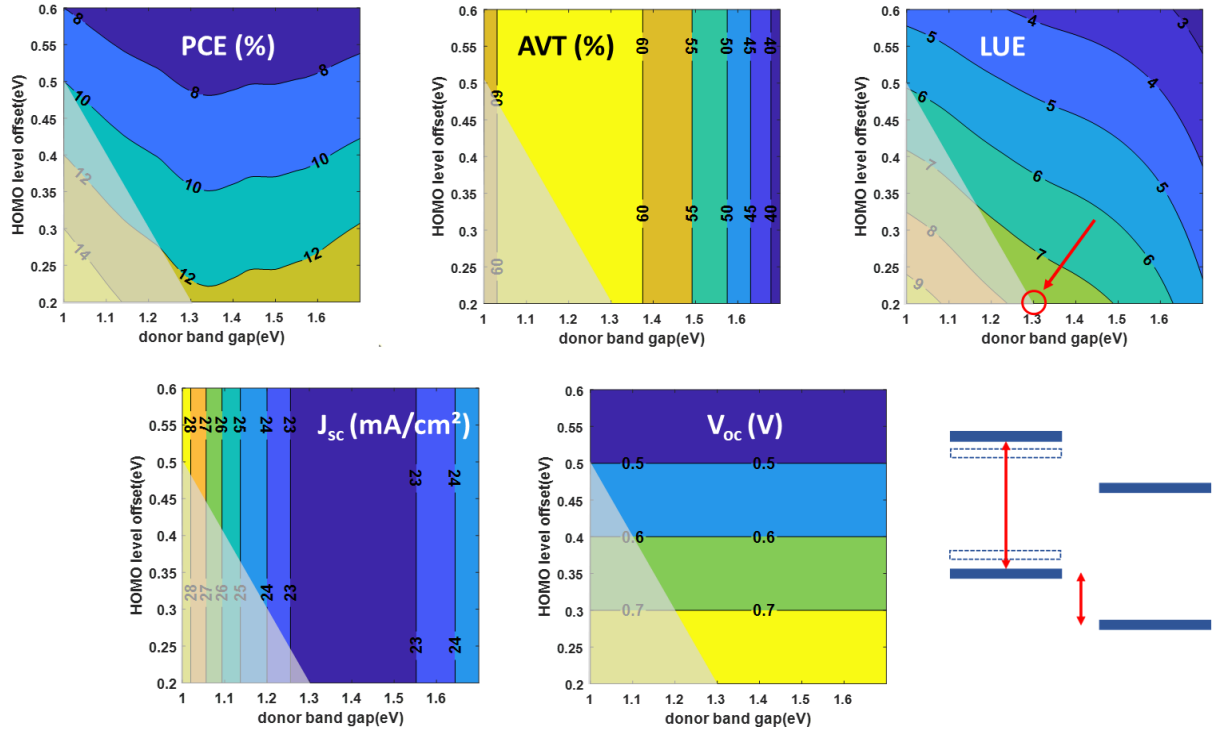

*Figure 2: PCE, AVT, LUE, $J_{sc}$ and $V_{oc}$ calculated for a variation of donor band gap and HOMO level offset, as indicated in the energy level schematic in the lower right. The energy levels of the acceptor were assumed as constant with $E_g = 1.3$ eV, and the donor energy levels as variable.*

Here, as in the other plots, the shaded grey areas indicate those input parameters for which the minimum energy level offset of 0.2 eV is no longer valid and a working solar cell cannot be guaranteed due to the absence of exciton splitting. For instance, the smallest donor band gap that ensures the minimum offset for $E_{g,\,acceptor} = 1.3$ eV also has a value of 1.3 eV. If the HOMO level offset is increased, the donor band gap can be smaller by the respective energy value, i.e., a HOMO level offset of 0.3 eV allows a donor band gap which is 0.1 eV smaller than the acceptor band gap, i.e., 1.2 eV as opposed to 1.3 eV.

From the simulation plots displayed in Figure 2, AVT, $J_{sc}$, and $V_{oc}$ show the expected behavior: As the acceptor absorption is almost completely outside of the visible range, AVT only depends on $E_{g,\,donor}$ and exhibits a broad maximum between 1.0 eV and 1.4 eV, because the absorption maximum is far from the visible region in this range. With increasing band gap, the AVT decreases, as the donor absorption moves into the visible region of the spectrum. $J_{sc}$ also only depends on the donor band gap and has a broad minimum between 1.3 eV and 1.5 eV where the absorption spectra of donor and acceptor overlap. When the overlap is decreased by shifting the donor absorption spectrum to the blue or further into the infrared, the total absorption increases, thus leading to slightly increased $J_{sc}$. As $V_{oc}$ does not change with the donor bandgap, PCE follows the course of $J_{sc}$. $V_{oc}$ only depends on the HOMO level offset and decreases linearly with it, due to the constant acceptor band gap. Accordingly, the resulting LUE has its maximum of 7.7 for the minimum energy level offset and the smallest donor band gap (i.e., for donor and acceptor band gaps being equal). This is intuitive since minimizing the energy level offset is a well-known design rule for OSCs to maximize $V_{oc}$. To confirm this conclusion, we repeated the calculation for three more values of $E_{g, acceptor}$, namely 1.1 eV, 1.6 eV, 1.9 eV, with the resulting LUE values displayed in Figure S2. In all cases, the highest LUE is obtained for equal band gaps of donor and acceptor. The absolute LUE values are lower for values of $E_{g,acceptor}$ either smaller or larger than 1.3 eV, which is due to lower voltage for 1.1 eV and due to decreased $J_{sc}$ and AVT for 1.6 eV and particularly for 1.9 eV. We therefore conclude that $E_g$ values are the most relevant parameters for device optimization, and that the energy level offset should be as close as possible to the minimum value. In the remainder of the manuscript, we therefore only vary the band gaps of the donor and the acceptor. The $E_g$ values that we consider range from 1.0 eV to 3.0 eV, with some of the plots zoomed in into the relevant range for better readability.

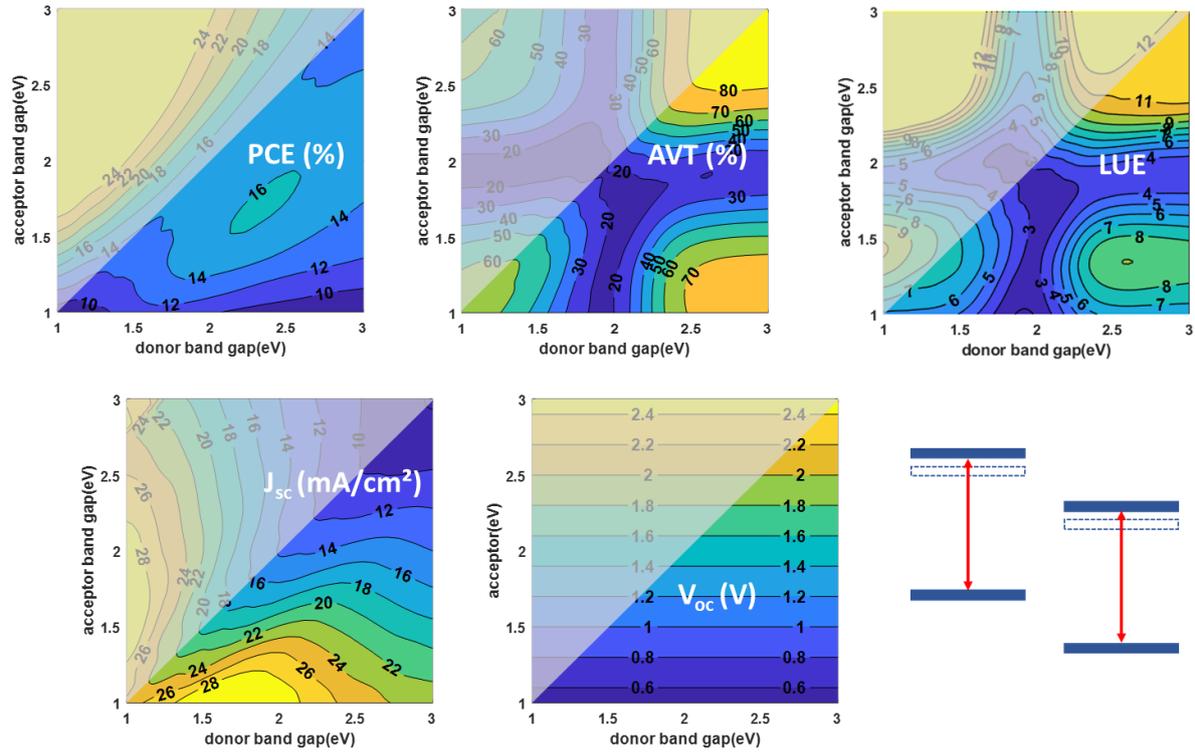

*Figure 3: PCE, AVT, LUE, $J_{sc}$ and $V_{oc}$ obtained for a variation of both, donor and acceptor band gaps. Both HOMO levels are constant, with an offset of 0.2 eV, whereas both LUMO levels are variable.*

Figure 3 displays the results obtained using the spectra of PCE10 and FOIC as donor and acceptor, respectively, and $OD_{max} = 0.7$. For this simulation, both HOMO level positions are assumed as constant with an offset of 0.2 eV, and both LUMO levels are assumed as variable. As a result, the condition for the minimum energy level offset is fulfilled when the donor band gap is equal to or larger than the acceptor band gap (area above the diagonal is shaded in the plots). As expected, $V_{oc}$ increases linearly with the acceptor band gap, as long as the condition for exciton splitting is fulfilled. $J_{sc}$ exhibits for equal band gaps (i.e. along the diagonal), due to the overlap of donor and acceptor absorption spectra, and broad maxima for complementary absorption belonging to the minimum band gap of 1 eV for the acceptor and ~1.4 – 1.9 eV for the donor, or vice versa. The AVT has maxima at the corners of the plot where donor and acceptor have the largest or the smallest band gaps, with the maximum value of over 80%, and its minimum in the middle where both donor and acceptor have a band gap of ~2 eV, corresponding to an absorption spectrum with the maximum overlap with the photopic curve. As a result, since the PCE shows less variation than the AVT, the LUE also has minima for band gaps of ~2 eV, and three distinct maxima in those regions where the band gaps are either smaller or larger, corresponding to reduced absorption in the visible region. The first maximum of LUE ~8 is found for identical band gaps of ~1.3 eV, the second maximum of LUE ~9 is found for a large donor band gap of ~2.6 eV and a smaller acceptor band gap of ~1.35 eV, and the third maximum of LUE ~12 is found for high band gap value of ~2.6 eV for both donor and acceptor, exhibiting by far the highest overall value due to the high $V_{oc}$.

These findings are not restricted to a certain material system or parameters, as can be seen in Figure 4 where the LUE is shown exemplarily for different values of $OD_{max}$, different donor: acceptor spectrum combinations and different donor: acceptor ratios (if not specified otherwise, the default value for $r_{D:A}$ is 1:1): in all cases, minimum and maximum values appear for almost the same band gaps as described above, only with different absolute values. In particular, the case depicted in Figure 3a, PCE10: FOIC spectra with $OD_{max} = 0.7$ (same parameters that were already used for the results shown in Figure 2) leads to the highest LUE values. A smaller $OD_{max}$ of 0.4 slightly reduces the LUE values, due to a stronger decrease in the PCE compared to the AVT. With respect to material variation, the exchange of the model spectra of PCE10: FOIC with those of PM6:Y6 leads to a decrease in LUE since the absorption spectrum of Y6 is broader than the one of FOIC and thus leads to enhanced absorption in the visible region (compare Figure 1).

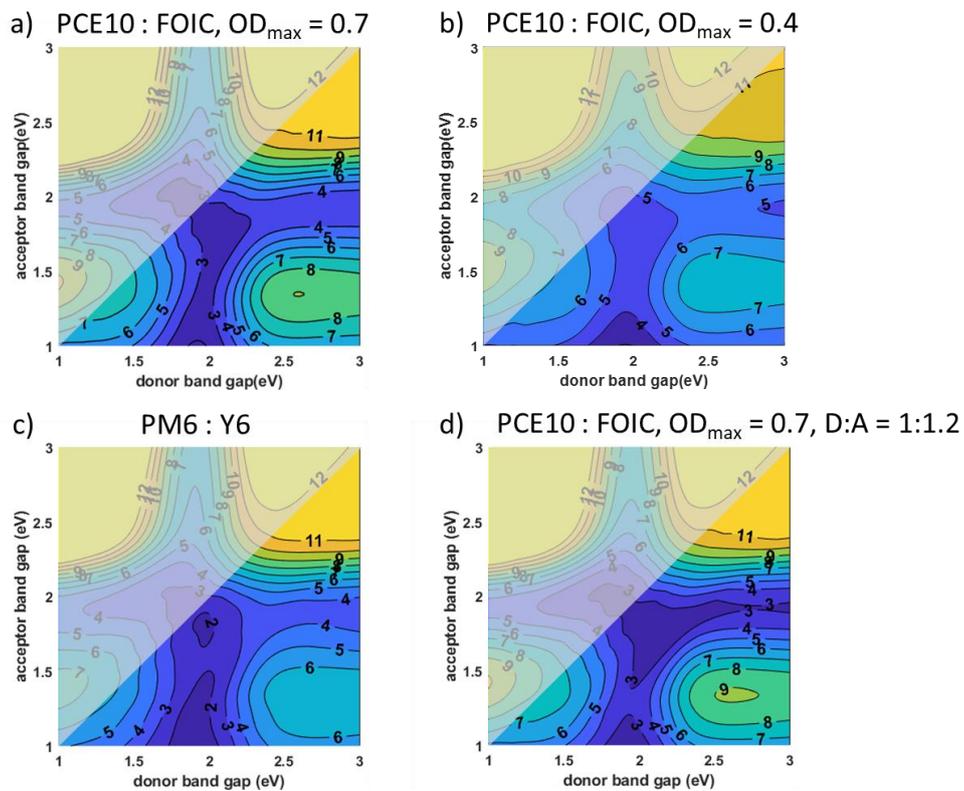

Figure 4: LUE calculated for absorption spectra corresponding to different donor: acceptor combinations, for different values of $OD_{max}$ (default value is 0.7), and for different D:A ratios.

*a) PCE10: FOIC with OD$_{max}$ = 0.7, b) PCE10: FOIC with OD$_{max}$ = 0.4, c) PM6:Y6, d) slightly increased acceptor content (D:A ratio of 1:1.2 for PCE10: FOIC and OD$_{max}$ = 0.7)*

After identifying the regions of high LUE, we compare the results of our model to values from the literature, both for publications showing record performance with respect to LUE as well as for material systems with band gap values that are reported less often. For record LUE, we analyzed data that is depicted in the PCE versus AVT plot on the Emerging PV website [3]. This comparison reveals the proximity of real materials and devices to their optimum performance that we aim to assess with our calculations, as shown in Figure 5. For the first row (a-c), we chose various publications with AVT values of ~50% [18-23]. The restriction to a certain AVT value was chosen because, as mentioned before, the simulation results always correspond to a certain OD$_{max}$ and consequently a certain transmittance. When analyzing the literature of OSCs with respect to the E$_g$ values of the absorber materials, it becomes obvious that all high-efficiency devices, both opaque and semitransparent, were obtained with a very narrow range of band gaps (see Figure S2 for an overview of the band gap values for all the publications with high LUE that are referenced in this manuscript). PM6 and PTB7-Th (1.8 eV and 1.6 eV according to our definition) or other, very similar, materials were used as donors almost exclusively (For opaque devices, see for instance the plot of PCE vs E$_g$ at the Emerging PV website, where all high efficiency OSC devices exhibit E$_{g, donor}$ values of 1.4 – 1.45 eV). While a larger range of acceptors with more structural variation has been employed, the band gaps of these materials are all in the range of 1.3eV – 1.4 eV. (The difference of 0.05 eV – 0.1 eV between the values used in paper and those given ot the Emerging PV website can be explained by the fact that the Emerging PV website uses the inflection point on the long wavelength side of the EQE [2]), whereas in this study we use the optical band gap as explained above.) In Figure 5a-c, the band gap range is limited to values between 1.2 eV and 1.4 eV for the acceptor and values between 1.55 eV and 1.85 eV for the donor for better legibility, as no appropriate data was found outside of this region. The literature data is shown as circles with the actual value represented by the color according to the color bar. For the exact values, please refer to Tables S1- S3. Since the majority of the publications corresponding to panels a-c utilized PCE10 as a donor, we selected the corresponding model absorption spectrum for the donor. With respect to the acceptor, several different ones were used in the respective publications. For the simulations, we chose the one of FOIC since it has the potential for highest LUE according to Figure 4.

Furthermore, the donor: acceptor ratio was set to 1:1.2, and $OD_{max}$ was chosen as 0.65 since this value leads to generally good agreement for the AVT values (Figure 5a).

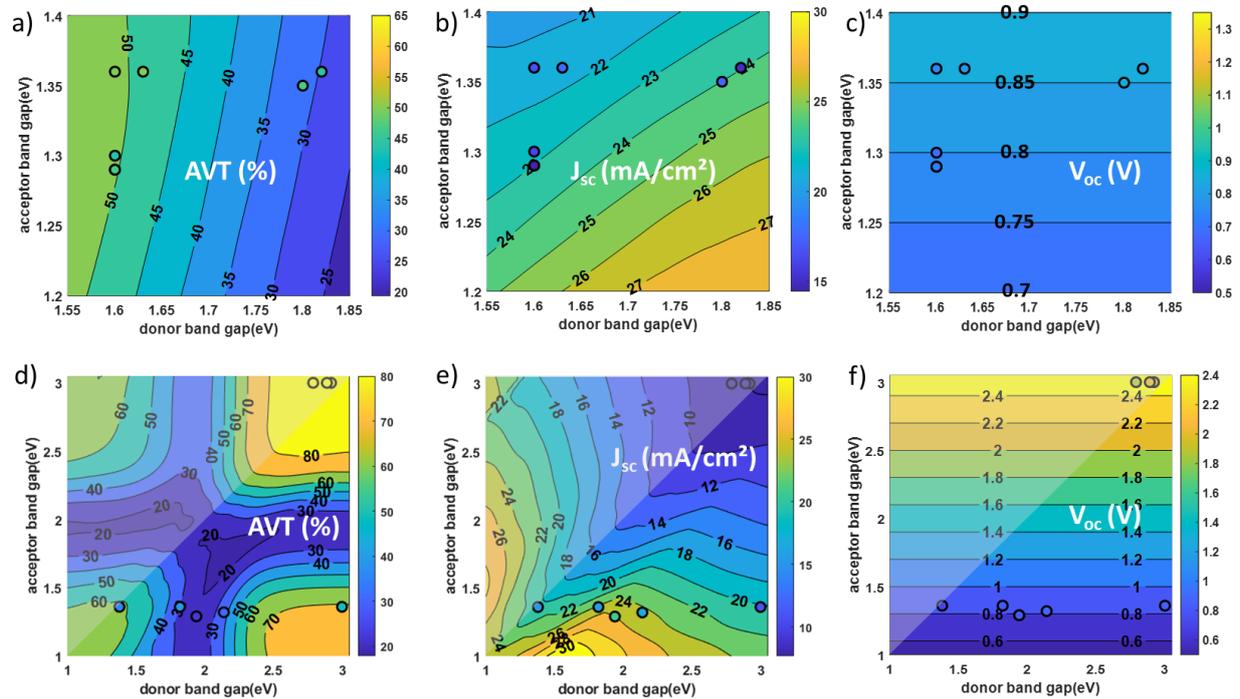

*Figure 5: Comparison of our model with data from the literature. a) – c): AVT, $J_{sc}$ and $V_{oc}$ for several publications with an AVT of ~50% and high LUE [18-23]. The calculation was performed with the absorption spectra of PCE10 and FOIC, $OD_{max}$ = 0.65 and $r_{DA}$ = 1:1.2. d)-f) AVT, $J_{sc}$ and $V_{oc}$ for a number of publications with uncommon band gaps [31-36]. The calculation was performed with the absorption spectra of PCE10 and FOIC, $OD_{max}$ = 0.65 and $r_{DA}$ = 1:1.2. For the data from [36], the nominal acceptor band gap of 3.44 eV has been shifted to 3.05 eV for the plot in order to stay within the previously used wavelength range.*

With respect to the agreement in $J_{sc}$, the calculated values lie between 21 mA/cm² and 24 mA/cm² whereas the measured ones are between 14.3 mA/cm² and 20.4 mA/cm². It has to be noted that our model, with the aim of providing a general description, neglects the influence of interference effects in the thin layer structure and it does not assume any outcoupling structure. Published results, on the other hand, often employ one or several dielectric layers on top of a thin metal electrode to increase AVT without compromising the $J_{sc}$, for instance the one with the highest $J_{sc}$ [19]. In general, after setting $OD_{max}$ to a value which reproduces the measured AVT, the experimental $J_{sc}$ values are lower than the calculated ones, which is likely because the real IQE values are lower than 90%. Another reason for the discrepancy might be due to the fact that the literature data was generated with different donors and acceptors, whereas the simulation always assumes the same absorption spectra. As mentioned before, the aim of this work is to provide a rationale for selecting or designing appropriate donor and acceptor materials by evaluating their potential, rather than making exact predictions of the expected efficiencies and AVTs.

Finally, the comparison of the calculated with the measured $V_{oc}$ values shows good agreement in general, with several measured values close to the calculated ones, especially for the devices made from PM6:Y6 (compare tables S1 – S3). This shows that the assumption of an additional voltage loss of 0.3 eV is consistent with our definition of the energy levels and the band gap. Experimental $V_{oc}$ values lower than the calculated ones are most likely due to non-optimized transport layers. Data from the literature with AVT values of ~20% [24-30] and high LUE are presented in Figure S3 leading to the same general conclusions as the analysis of higher AVT values shown in Figure 5. In this group of data, there is one point that exhibits worse agreement of the AVT ([30], 30% predicted *vs* 20%); this difference might come from the fact that it is the only one for which the OSC device was not fabricated with PM6:Y6, but with PTB7-th: ATT-9. However, the more likely reason is the unusually thick silver electrode (18 nm instead of the commonly used 10 nm) which generally leads to an increase in $J_{sc}$ at the cost of reduced AVT.

Since, as already mentioned, the experimental data shown in Figures 5a-c are limited to a small number of band gap values, we also conduct a targeted search for publications with different $E_g$ values of donor or acceptor and display them in the lower row of Figure 5. In this group of data, there are some publications with donor band gaps larger than 1.8 eV [31-35], and one publication [36] in which both donor and acceptor band gaps are around 3 eV. According to our search, not many OSCs with different band gaps have been published. In general, the agreement is worse for this group of publications than for the one in the top row because they did not use optimized systems. For instance, the measured AVT of 35.8% for the combination of low band gap donor PM2 with Y6:BO ([35], located in the lower left corner at 1.38 eV / 1.36 eV in Figure 5d) is low compared to the calculated AVT value of over 60%. Similarly, the value for the high band gap donor FC-S1:PM6 combined with Y6:BO ([32], shown in the lower left corner at 3.02 eV / 1.36 eV) was measured as 49.3 % whereas the model predicts over 70 %. This discrepancy can be explained by the fact that Y6-BO has a broader absorption spectrum than the used model spectrum of FOIC, and that a certain percentage of PM6 was added to the blend to ensure a well-working solar cell. For the devices combining the high band gap donors BF-DPB, BF-DPT and BF-DPN with the high band gap acceptor B4PymPM [36], the PCE values are below 1% due to $J_{sc}$ values below 1 mA/cm². However, the AVT values are above 80 %, and the $V_{oc}$ values lie around 2V, approaching the predicted values of 2.5 eV.

Up to now, we have used the photopic curve, the sensitivity curve of the human eye to quantify the transparent property of the solar cells with the AVT. This figure of merit is particularly suitable for windows in building – integrated applications. In addition, our model can be easily adapted to other applications such as greenhouses by simply replacing the photopic curve with the plant action spectrum [38], which exhibits two maxima at 450 nm and 670 nm (Figure S4). Analogous to the AVT, we can now define the APART as

$$APART = \frac{\int T(\lambda) * PAR(\lambda) * \Gamma_{AM1.5G}(\lambda)d\lambda}{\int P(\lambda) * \Gamma_{AM1.5G}(\lambda)d\lambda}$$

and the corresponding light utilization efficiency as

$$LUE^* = PCE * APART.$$

The results of this calculation are shown in Figure 6 for APART and LUE*, for a slightly reduced value of $OD_{max}$ = 0.5. PCE, $J_{sc}$ and $V_{oc}$ are not shown because they do not change compared to the results shown in Figure 2.

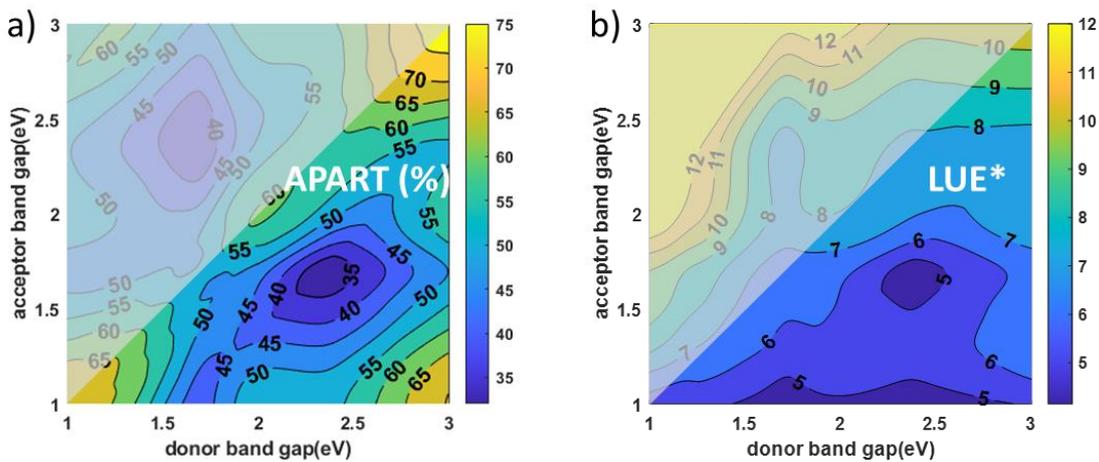

*Figure 6: APART and LUE obtained for a variation of both, donor and acceptor band gaps. Both HOMO levels are constant, with an offset of 0.2 eV, whereas both LUMO levels are variable. $J_{sc}$, $V_{oc}$ and PCE are the same as in Figure 3.*

In this case, since the ideal transmittance spectrum of the solar cell has two maxima, the first one in the blue region around 450 nm and the second one in the red region around 670 nm, the APART exhibits a maximum instead of a minimum value of ~60%, for band gap values of 2 eV, resulting in a PCE of around 16%. The resulting LUE* exhibits an according maximum of

~8 at band gap values of ~2 eV. Compared to the LUE plots in Figures 2 and 3, the maximum appearing at smaller band gaps has been shifted to values of 1.3 eV / 1.3 eV and is lower than the one appearing at 2 eV / 2 eV. The maximum found previously for large donor band gap combined with small acceptor band gap has disappeared, whereas the maximum for large donor band gap / large acceptor band gaps still exists, but is not as pronounced anymore compared to the other maxima.

Finally, we want to estimate how realistic it is to synthesize materials with a certain band gap, to achieve the high LUE values predicted by our model. For this purpose, we consider a recent publicly available database [39] of orbital energy levels of known molecular semiconductor and their associated (optical) band-gap which was computed from quantum chemical methods for ~50k molecular semiconductors whose solid structure is deposited in the Cambridge Structural Database.[40] While this set is by no means comprehensive, it can be considered an unbiased sample of the chemical space and it has been used to discover lead compounds for different optoelectronic applications.[41] For all possible molecular pairs that can be constructed, we considered only type-II junctions, i.e., those with a favorable alignment of orbital energies of $\Delta E_{HOMO}$ and $\Delta E_{LUMO}$ in the 0.2-0.4 eV range. In line with the previous analysis, we focused on the ~700k pairs of donors and acceptors constructed from the database with $E_{g,donor}$ and $E_{g,acceptor}$ in the 1-3 eV range. Figure 7 reports the number of pairs found in different regions of the donor and acceptor gap in bins of 0.2×0.2 eV. Noting the log scale of the color map, one can immediately find that the vast majority of pairs forming type-II junctions are found for large gaps where tens of thousands of potential pairs can be considered. However, a few tens of instances are also found in the maximum of LUE with donor and acceptor band gaps around 1.4 eV and a few hundred in the other LUE maximum with small band gap of acceptor and large band gap of donor. With this approach we can confirm that all LUE maxima are chemically feasible, with the one at low band gaps being comparatively more challenging to realize.

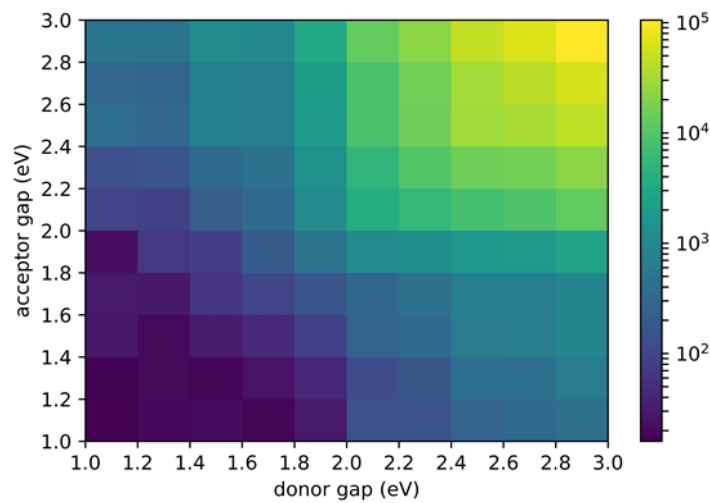

**Figure 7**. *Number of donor-acceptors pairs extracted from the database of computed molecular properties [39] with favorable alignment of donor and acceptor HOMO and LUMO level and with different values of donor and acceptor gap.*

**Conclusion**

We have performed a study to determine the optimum performance of semitransparent organic solar cells based on the band gaps of the donor and acceptor materials. The aim of the work is to provide a rationale and a guideline for selecting or designing appropriate donor and acceptor materials, rather than making exact predictions of the expected PCEs and AVTs. Our model is based on the model by Scharber et al.[7], incorporating actual absorbance spectra of commonly used materials for semitransparent cells that are adjusted in energy to accommodate variable band gaps.

Independent of the particular assumptions that are made for material spectra, optical density etc., the results show that the maximum LUE is found for a minimum energy level offset, and in three distinct band gap regions: firstly, for small donor and acceptor band gaps of ~1.4 eV, secondly for a small acceptor band gap of ~1.4 eV combined with a large donor band gap of ~2.6 eV, and thirdly for large donor and acceptor band gaps of ~2.6 eV. The comparison with literature data reveals that all high-performing semitransparent OSCs have been demonstrated with two different donors with band gaps of 1.6 eV and 1.8 eV respectively, and a variety of acceptors with band gap values between 1.3 and 1.4 eV. Our analysis shows that current record devices with respect to LUE do not employ materials with optimized band gaps, and devices with more suitable band gaps are not optimized with respect to photovoltaic performance.

Our model can be easily adapted to other applications by replacing the weighting spectrum, for instance with the plant action spectrum, and thus making it suitable to assess the potential of a certain material for greenhouse applications. The result shows that, in this case, optimum performance can be expected for both donor and acceptor band gaps of 2.0 eV.

Finally, we analyze the availability of suitable molecules with a database search and find that all band gap values are chemically feasible. We therefore encourage the community to synthesize both donors and acceptors with optimized absorption spectra and favorable transport properties to fully exploit the application potential of semitransparent OSCs.

**Acknowledgments**

A.T. acknowledges the support of the European Innovation Council, Project No. 101057564. K.F. and C.L. acknowledge the support of the H2020 European project Citysolar, grant agreement number 101007084. The authors acknowledge the 'Solar Factory of the Future' as part of the Energy Campus Nuremberg (EnCN), which is supported by the Bavarian State Government (FKZ 20.2-3410.5-4-5). H.-J. E. and C.J.B. acknowledge funding from the European Union's Horizon 2020 INFRAIA program under Grant Agreement No. 101008701 ('EMERGE'). Part of this work has been supported by the Helmholtz Association in the framework of the innovation platform "Solar TAP".

Larry Lüer, Andreas Distler and Iain McCulloch are acknowledged for very helpful discussions. Fenzhi Ai is acknowledged for her help with the literature review.

Supporting Information

# Guidelines for material design for semitransparent organic solar cells

Karen Forberich[1,2], Alessandro Troisi[3], Chao Liu[2], Michael Wagner[1], Christoph J. Brabec[1,2], Hans – Joachim Egelhaaf[1,2]

*[1]Forschungszentrum Jülich GmbH, Helmholtz Institute Erlangen-Nürnberg for Renewable Energy (HI ERN), Dept. of High Throughput Methods in Photovoltaics, Erlangen, Germany*

*[2]Friedrich-Alexander-Universität Erlangen-Nürnberg, Materials for Electronics and Energy Technology (i-MEET), Erlangen, Germany*

[3]*Dept. of Chemistry, University of Liverpool, United Kingdom*

*k.forberich@fz-juelich.de*

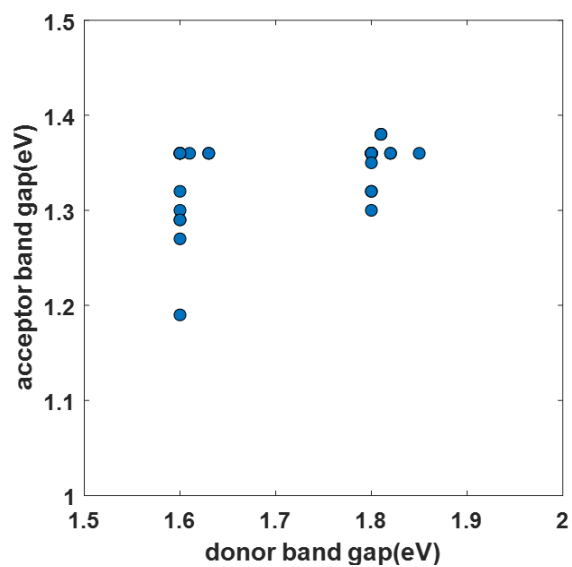

*Figure S1: Overview of band gap values of all the references cited in this manuscript, with band gaps determined according to the definition in Figure 1b. References that are not mentioned in the main manuscript are listed at the end of the supporting information.*

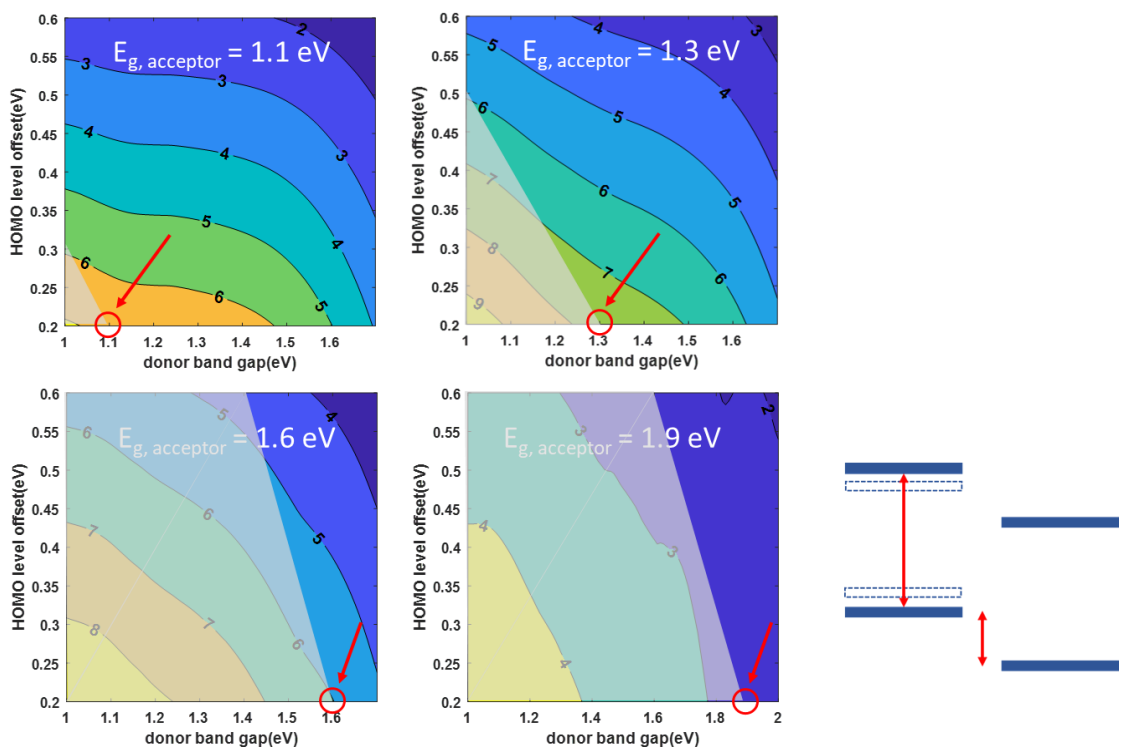

*Figure S2: LUE calculated for a variation of donor band gap and HOMO level offset, analogous to Figure S2, for four different band gap values of the acceptor. The highest LUE value (6.7, 7.7, 6.0, and 2.9 for $E_g$ of 1.1 eV, 1.3 eV, 1.6 eV, and 1.9eV, respectively) is always obtained for equal band gaps of donor and acceptor, i.e. for the minimum HOMO level offset of 0.2 eV and the smallest allowed band gap.*

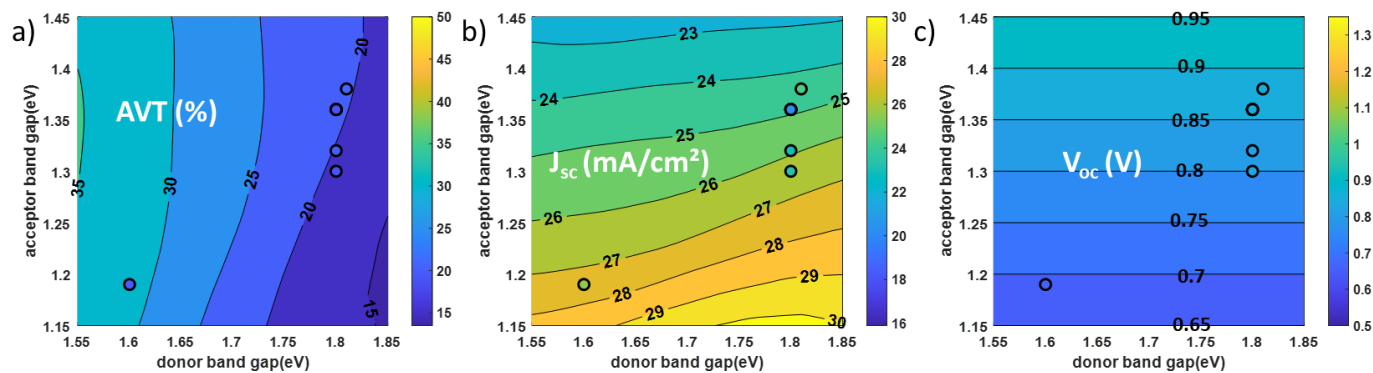

*Figure S3: Comparison of our model with data from the literature. a) – c): AVT, $J_{sc}$ and $V_{oc}$ for several publications with an AVT of ~20% [24-30]. The calculation was performed with the absorption spectra of PM6 and Y6, $OD_{max} = 0.65$ and $r_{DA} = 1:1.2$.*

| Ref | ePV ID | Donor | Acceptor | $E_{g,donor}$ (eV) | $E_{g,acceptor}$ (eV) | $PCE_{measured}$ (%) | $PCE_{calculated}$ (%) | $V_{oc,measured}$ (V) | $V_{oc,calculated}$ (V) | $FF_{measured}$ (%) |
|---|---|---|---|---|---|---|---|---|---|---|
| 18 | 615 | PM6 | BTP-eC9:L8-BO | 1.8 | 1.35 | 11.1 | 14.4 | 0.852 | 0.85 | 74.1 |
| 19 | 338 | PCE10 | A078 | 1.6 | 1.29 | 10.8 | 12.8 | 0.75 | 0.79 | 70.6 |
| 20 | 645 | PCE10-2F | Y6 | 1.63 | 1.36 | 10.01 | 13.1 | 0.79 | 0.86 | 71 |
| 21 | 342 | PTB7-Th | FOIC:PC71BM | 1.6 | 1.36 | 8.3 | 12.9 | 0.746 | 0.86 | 66.8 |
| 22 | 796 | PBT1-C-2C | Y6 | 1.82 | 1.36 | 8.2 | 14.5 | 0.806 | 0.86 | 67.6 |
| 23 | 337 | PCE-10 | BT-CIC:TT-FIC | 1.6 | 1.3 | 8 | 12.8 | 0.68 | 0.8 | 72.6 |
| 19 | 339 | PCE10 | A078 | 1.6 | 1.29 | 7.1 | 12.8 | 0.73 | 0.79 | 68 |

| Ref. | ePV ID | Donor | Acceptor | $E_{g,donor}$ | $E_{g,acceptor}$ | $J_{sc,measured}$ (mA/cm²) | $J_{sc,calculated}$ (mA/cm²) | AVT measured (%) | AVT calculated (%) |
|---|---|---|---|---|---|---|---|---|---|
| 18 | 615 | PM6 | BTP-eC9:L8-BO | 1.8 | 1.35 | 17.57 | 24.3 | 47 | 31.8 |
| 19 | 338 | PCE10 | A078 | 1.6 | 1.29 | 20.4 | 23.1 | 46 | 50.1 |
| 20 | 645 | PCE10-2F | Y6 | 1.63 | 1.36 | 17.8 | 21.7 | 50 | 49.5 |
| 21 | 342 | PTB7-Th | FOIC:PC71BM | 1.6 | 1.36 | 16.7 | 21.5 | 50 | 51.6 |
| 22 | 796 | PBT1-C-2C | Y6 | 1.82 | 1.36 | 15.1 | 24.3 | 44.2 | 30.0 |
| 23 | 337 | PCE-10 | BT-CIC:TT-FIC | 1.6 | 1.3 | 16.2 | 22.9 | 44 | 51.2 |
| 19 | 339 | PCE10 | A078 | 1.6 | 1.29 | 14.3 | 23.1 | 47 | 50.1 |

*Table S1: Reference and properties for the literature data with AVT ~50% that is shown in Figure 5 a-c), with simulated values given as comparison. The fill factor assumed for the calculation is always 70%.*

| Ref. | ePV ID | Donor | Acceptor | $E_{g,donor}$ | $E_{g,acceptor}$ | PCE$_{measured}$ (%) | PCE$_{calculated}$ (%) | $V_{oc, measured}$ (V) | $V_{oc, calculated}$ (V) | FF$_{measured}$ (%) |
|---|---|---|---|---|---|---|---|---|---|---|
| 24 | 619 | PM6-Ir1 | BTP-ec9:PC71BM | 1.81 | 1.38 | 16.1 | 15.1 | 0.86 | 0.88 | 76.1 |
| 25 | 617 | PM6 | ICBA:Y6 | 1.8 | 1.36 | 14.6 | 15.0 | 0.86 | 0.86 | 74.7 |
| 26 | 622 | PM6 | Y7 | 1.8 | 1.3 | 13.63 | 14.8 | 0.83 | 0.8 | 69.8 |
| 27 | 621 | PM6 | Y6:SN | 1.8 | 1.32 | 14 | 14.9 | 0.82 | 0.83 | 73.8 |
| 28 | 323 | PBDB-T-2F | Y6 | 1.8 | 1.36 | 11.7 | 15.0 | 0.81 | 0.86 | 69.6 |
| 29 | 324 | PM6 | Y6 | 1.8 | 1.36 | 12.4 | 15.0 | 0.852 | 0.86 | 71.4 |
| 30 | 529 | PTB7-th | ATT-9 | 1.6 | 1.19 | 11.56 | 13.3 | 0.661 | 0.69 | 68.2 |
| 28 | 325 | PBDB-T-2F | Y6 | 1.8 | 1.36 | 10.5 | 15.0 | 0.8 | 0.86 | 68.3 |
| 28 | 322 | PBDB-T-2F | Y6 | 1.8 | 1.36 | 12.6 | 15.0 | 0.81 | 0.86 | 73.2 |

| Ref. | ePV ID | Donor | Acceptor | $E_{g,donor}$ | $E_{g,acceptor}$ | $J_{sc, measured}$ (mA/cm²) | $J_{sc, calculated}$ (mA/cm²) | AVT measured (%) | AVT calculated (%) |
|---|---|---|---|---|---|---|---|---|---|
| 24 | 619 | PM6-Ir1 | BTP-ec9:PC71BM | 1.81 | 1.38 | 24.6 | 24.4 | 21 | 20.1 |
| 25 | 617 | PM6 | ICBA:Y6 | 1.8 | 1.36 | 22.75 | 24.9 | 20 | 20.7 |
| 26 | 622 | PM6 | Y7 | 1.8 | 1.3 | 23.4 | 26.4 | 19 | 19.1 |
| 27 | 621 | PM6 | Y6:SN | 1.8 | 1.32 | 23 | 25.9 | 20 | 19.8 |
| 28 | 323 | PBDB-T-2F | Y6 | 1.8 | 1.36 | 20.7 | 24.9 | 18 | 20.7 |
| 29 | 324 | PM6 | Y6 | 1.8 | 1.36 | 20.4 | 24.9 | 19 | 20.7 |
| 30 | 529 | PTB7-th | ATT-9 | 1.6 | 1.19 | 25.63 | 27.4 | 20 | 31.5 |
| 28 | 325 | PBDB-T-2F | Y6 | 1.8 | 1.36 | 19.3 | 24.9 | 21 | 20.7 |
| 28 | 322 | PBDB-T-2F | Y6 | 1.8 | 1.36 | 21.2 | 24.9 | 17 | 20.7 |

*Table S2: Reference and properties for the literature data with AVT ~20% that is shown in Figure S2 a-c). The fill factor assumed for the calculation is always 70%.*

| Ref. | ePV ID | Donor | Acceptor | $E_{g,donor}$ | $E_{g,acceptor}$ | $PCE_{measured}$ (%) | $PCE_{calculated}$ (%) | $V_{oc,measured}$ (V) | $V_{oc,calculated}$ (V) | $FF_{measured}$ (%) |
|---|---|---|---|---|---|---|---|---|---|---|
| 31 | 793 | PBOF | eC9:L8-BO | 2.14 | 1.32 | 10.01 | 14.1 | 0.88 | 0.82 | 68.0 |
| 36 | N/A | BF-DPB | B4PymPM | 2.92 | 3.44 | 0.7 | 14.1 | 2 | 2.55 | 55.0 |
| 36 | N/A | BF-DPT | B4PymPM | 2.89 | 3.44 | 0.43 | 14.2 | 2.03 | 2.55 | 40.0 |
| 36 | N/A | BF-DPN | B4PymPM | 2.79 | 3.44 | 0.61 | 14.8 | 1.88 | 2.55 | 43.0 |
| 32 | 788 | FC-S1:PM6 | Y6:BO | 3.02 | 1.36 | 6.01 | 12.1 | 0.851 | 0.86 | 63.8 |
| 33 | N/A | D18 | N3 | 1.94 | 1.29 | 12.58 | 14.5 | 0.836 | 0.79 | 72.1 |
| 34 | N/A | D18 | N3 | 1.94 | 1.29 | 12.9 | 14.5 | 0.831 | 0.79 | 74.4 |
| 22 | 796,797 | PBT1-C-2Cl | Y6 | 1.82 | 1.36 | 9.1 | 14.5 | 0.82 | 0.86 | 67.2 |
| 35 | N/A | PM2 | Y6:BO | 1.38 | 1.36 | 7.5 | 12.4 | 0.71 | 0.86 | 68.7 |

| Ref. | ePV ID | Donor | Acceptor | $E_{g,donor}$ | $E_{g,acceptor}$ | $J_{sc,measured}$ (mA/cm²) | $J_{sc,calculated}$ (mA/cm²) | AVT measured (%) | AVT calculated (%) |
|---|---|---|---|---|---|---|---|---|---|
| 31 | 793 | PBOF | eC9:L8-BO | 2.14 | 1.32 | 16.82 | 24.5 | 30.48 | 36.2 |
| 36 | N/A | BF-DPB | B4PymPM | 2.92 | 3.44 | 0.64 | 8.0 | 81.8 | 84.9 |
| 36 | N/A | BF-DPT | B4PymPM | 2.89 | 3.44 | 0.53 | 8.0 | 82 | 84.8 |
| 36 | N/A | BF-DPN | B4PymPM | 2.79 | 3.44 | 0.76 | 8.4 | 80.3 | 84.6 |
| 32 | 788 | FC-S1:PM6 | Y6:BO | 3.02 | 1.36 | 11.07 | 20.2 | 49.28 | 71.1 |
| 33 | N/A | D18 | N3 | 1.94 | 1.29 | 20.87 | 26.0 | 22.81 | 23.6 |
| 34 | N/A | D18 | N3 | 1.94 | 1.29 | 20.9 | 26.0 | 22.49 | 23.6 |
| 22 | 796,797 | PBT1-C-2Cl | Y6 | 1.82 | 1.36 | 15.71 | 24.3 | 40.1 | 30.0 |
| 35 | N/A | PM2 | Y6:BO | 1.38 | 1.36 | 15.4 | 20.5 | 35.8 | 61.1 |

*Table S3: Reference and properties for the literature data that is shown in Figure 5 d-f).*

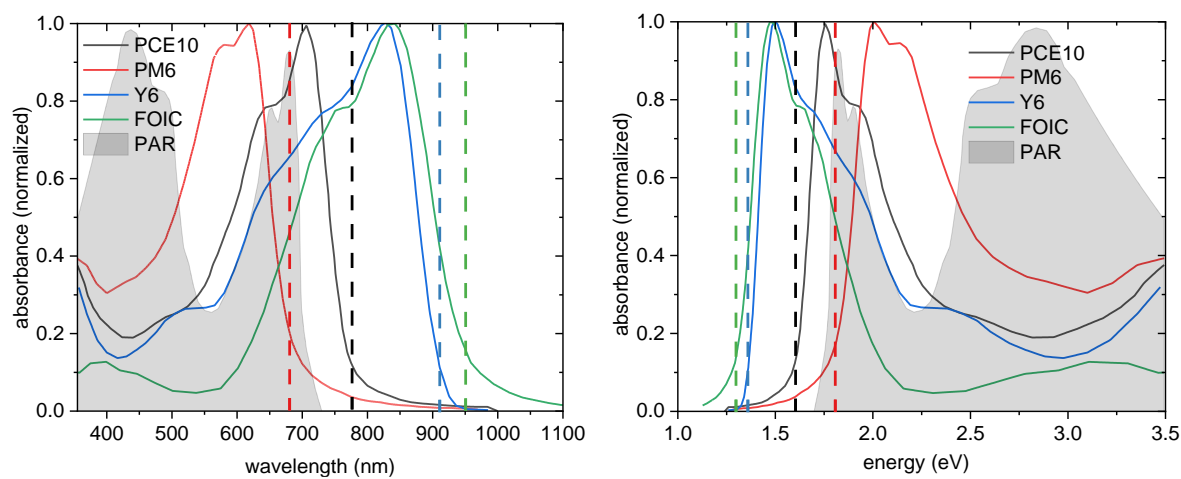

*Figure S4: Plant action spectrum, shown with the absorbance spectra used in this work over wavelength (left) and energy (right).*

**Additional References:**